\documentclass[USenglish,twocolumn]{article}

\usepackage[utf8]{inputenc}
\usepackage[big]{dgruyter}

\begin{document}

  \articletype{Research Article{\hfill}Open Access}

  \author*[1]{Stephan Geier}
  \author[2]{Roy H. \O stensen}
  \author[3]{Peter Nemeth}
  \author[3]{Ulrich Heber}
  \author[4]{Nicola P. Gentile Fusillo}
  \author[4]{Boris T. G\"ansicke}
  \author[5]{John H. Telting}
  \author[6]{Elizabeth M. Green}
  \author[3]{Johannes Schaffenroth}

  \affil[1]{Institute for Astronomy and Astrophysics, Kepler Center for Astro and Particle Physics, Eberhard Karls University, Sand 1, D 72076 T\"ubingen, Germany, E-mail: geier@astro.uni-tuebingen.de}
  \affil[2]{Department of Physics, Astronomy, and Materials Science, Missouri State University, Springfield, MO 65804, USA}
  \affil[3]{Dr.~Karl~Remeis-Observatory \& ECAP, Astronomical Institute, Friedrich-Alexander University Erlangen-Nuremberg, Sternwartstr.~7, D 96049 Bamberg, Germany}
  \affil[4]{Department of Physics, University of Warwick, Coventry CV4 7AL, UK}
  \affil[5]{Nordic Optical Telescope, Rambla Jos\'e Ana Fern\'andez P\'erez 7, E-38711 Brena Baja, Spain}
  \affil[6]{Steward Observatory, University of Arizona, 933 North Cherry Avenue, Tucson, AZ 85721, USA}
  
  \title{\huge Meet the family $-$ the catalog of known hot subdwarf stars}

  \runningtitle{Catalog of hot subdwarfs}


\begin{abstract}
{In preparation for the upcoming all-sky data releases of the Gaia mission, we compiled a catalog of
known hot subdwarf stars and candidates drawn from the literature and yet unpublished databases. The
catalog contains 5613 unique sources and provides multi-band photometry from the ultraviolet to the far
infrared, ground based proper motions, classifications based on spectroscopy and colors, published
atmospheric parameters, radial velocities and light curve variability information. Using several different
techniques, we removed outliers and misclassified objects. By matching this catalog with astrometric and
photometric data from the Gaia mission, we will develop selection criteria to construct a homogeneous,
magnitude-limited all-sky catalog of hot subdwarf stars based on Gaia data.\\
As first application of the catalog data, we present the quantitative spectral analysis of 280 sdB and sdOB stars from the Sloan Digital Sky Survey Data Release 7. Combining our derived parameters with state-of-the-art proper motions, we performed a full kinematic analysis of our sample. This allowed us to separate the first significantly large sample of 78 sdBs and sdOBs belonging to the Galactic halo. Comparing the properties of hot subdwarfs from the disk and the halo with hot subdwarf samples from the globular clusters $\omega$~Cen and NGC\,2808, we found the fraction of intermediate He-sdOBs in the field halo population to be significantly smaller than in the globular clusters.}
\end{abstract}

  \journalname{Open Astronomy}
\DOI{DOI}
  \startpage{1}
  \received{..}
  \revised{..}
  \accepted{..}

  \journalyear{2014}
  \journalvolume{1}

\maketitle
\section{Introduction}

Hot subdwarf stars (sdO/Bs) have spectra similar to main sequence O/B stars, but are subluminous and more compact. In the Hertzsprung-Russell diagram, those stars are located at the Extreme or Extended Horizontal Branch \citep[EHB,][]{heber86} and considered to be core helium-burning stars. To end up on the EHB, stars have to lose almost their entire hydrogen envelopes in the red-giant phase most likely via binary mass transfer. For a comprehensive review of the state-of-the-art hot subdwarf research see \citep{heber16}.

SdO/B stars were initially found in surveys looking for faint blue stars at high Galactic latitudes \citep{humason47}. \citet{kilkenny88} published the first catalog of spectroscopically identified hot subdwarf stars. Many more hot subdwarfs have been detected subsequently. \citet{oestensen06} created the hot subdwarf database containing more than 2300 stars. 

However, since 2006 the number of known hot subdwarfs again increased by a factor of more than two. The Sloan Digital Sky Survey (SDSS) provided spectra of almost 2000 sdO/Bs \citep{geier15,kepler15,kepler16}, reaching down to much fainter magnitudes than previous surveys. On the bright end of the magnitude distribution, new samples of hot subdwarfs have been selected from the EC survey and the GALEX all-sky survey photometry in the UV \citep[e.g.][]{kilkenny16,vennes11}. Furthermore, new large-area photometric and astrometric surveys are currently conducted in multiple bands from the UV to the far infrared. Given this wealth of new high quality data, we consider it timely to compile a new catalog of hot subdwarf stars \citep{geier17}. 

This catalog will be used as input and calibration dataset to select a magnitude-limited, homogeneous, all-sky catalog of hot subdwarf stars using astrometry and photometry from the Gaia mission, which will allow us to study the properties of the hot subdwarf population with unprecedented accuracy.

\section{Catalog content}

The catalog contains 5613 unique sources and provides multi-band photometry from the ultraviolet to the far infrared, ground based proper motions, classifications based on spectroscopy and colors, published atmospheric parameters, radial velocities and light curve variability information. Using several different techniques, we removed outliers and misclassified objects.
For the details of the catalog compilation see \citet{geier17}. The catalog is accessible online in the Vizier database.\footnote{Vizier catalog J/A+A/600/A50}

The next and yet unpublished version of the catalog contains the most recent ground-based proper motion catalogs combining their data with accurate positions from the Gaia mission Data Release 1: UCAC5, \citep{zacharias17}; HSOY, \citep{altmann17}; GPS1, \citep{tian17}. It also contains multi-band photometry from the PanSTARRS PS1 catalog.\footnote{https://archive.stsci.edu/panstarrs/} 

\section{Applying the catalog - population study of sdBs}

Detailed conclusions about possible formation scenarios of sdBs are difficult to draw. The unclear Galactic population membership of the sdB and sdOB stars is an important obstacle. Since most sdB stars are regarded as the progeny of low-mass stars, their progenitors can be quite old and therefore in principle belong to all Galactic populations, either the young thin disk, the older thick disk, or even the Galactic halo, which constitutes the oldest population. The main difference between these populations is the primordial metallicity, which influences the evolutionary histories of the sdB progenitor stars significantly. But also the age of the progenitor population can be an important factor for sdB formation.

Chemical tagging is normally used to assign stars to their Galactic populations. Metal-rich stars belong to the young population I, while metal-poor stars belong to the old population II. Because the abundances in sdB atmospheres are altered by diffusion processes, this quite straightforward method cannot be applied. Instead, the relatively close-by samples of field sdBs can be distinguished by their kinematical properties. While disk stars orbit the Galactic centre with moderate velocities and eccentricities, halo stars can have retrograde, very eccentric and highly inclined trajectories with respect to the Galactic plane. 

This method has been used to determine the population membership of rather bright samples of field sdBs \citep{deboer97,altmann04,kawka15,heber16,martin17}. All those studies concluded that most field sdBs are members of the Galactic disk and only few halo candidates have been found so far. 

\begin{figure*}[t!]
\begin{center}
	\resizebox{8cm}{!}{\includegraphics{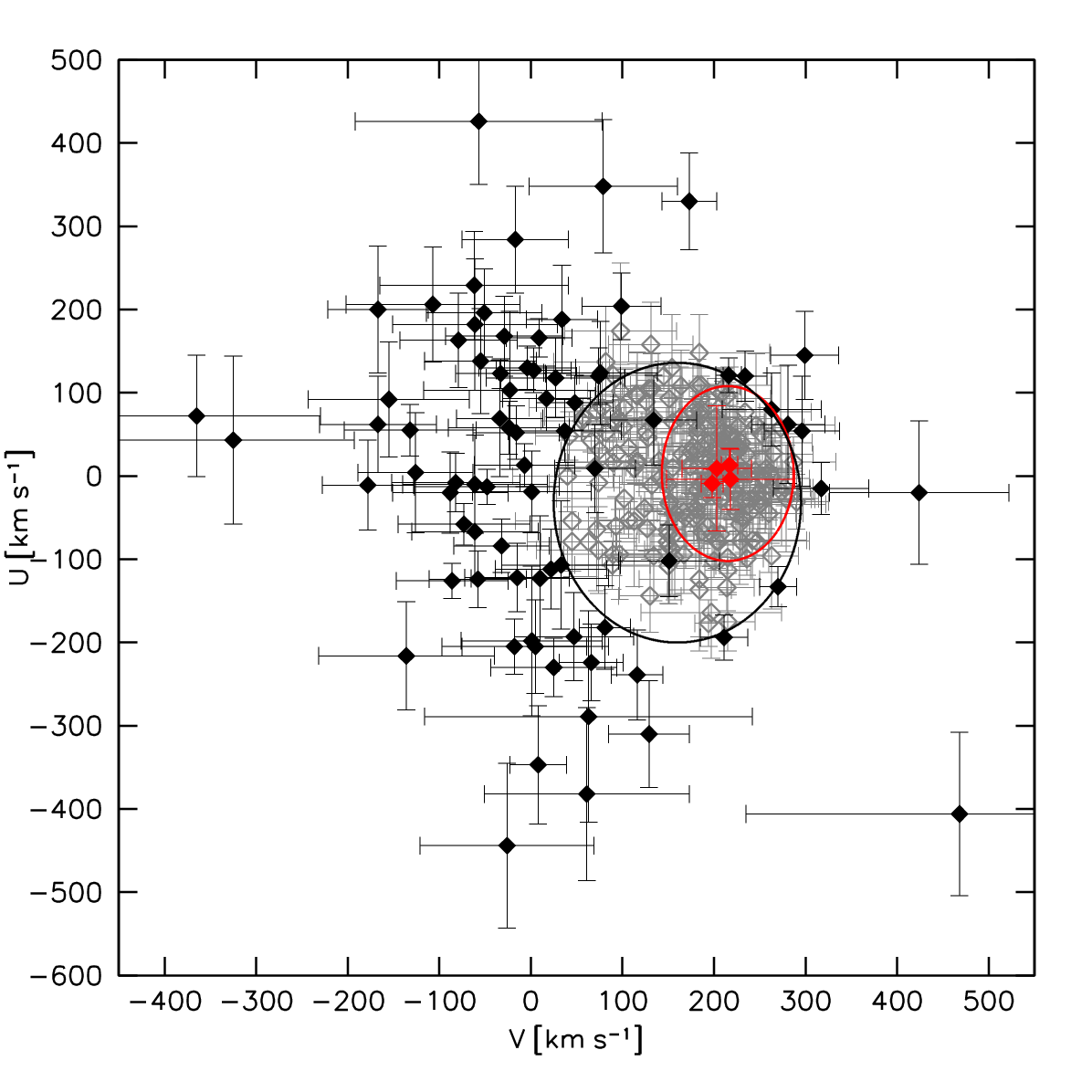}}
	\resizebox{8cm}{!}{\includegraphics{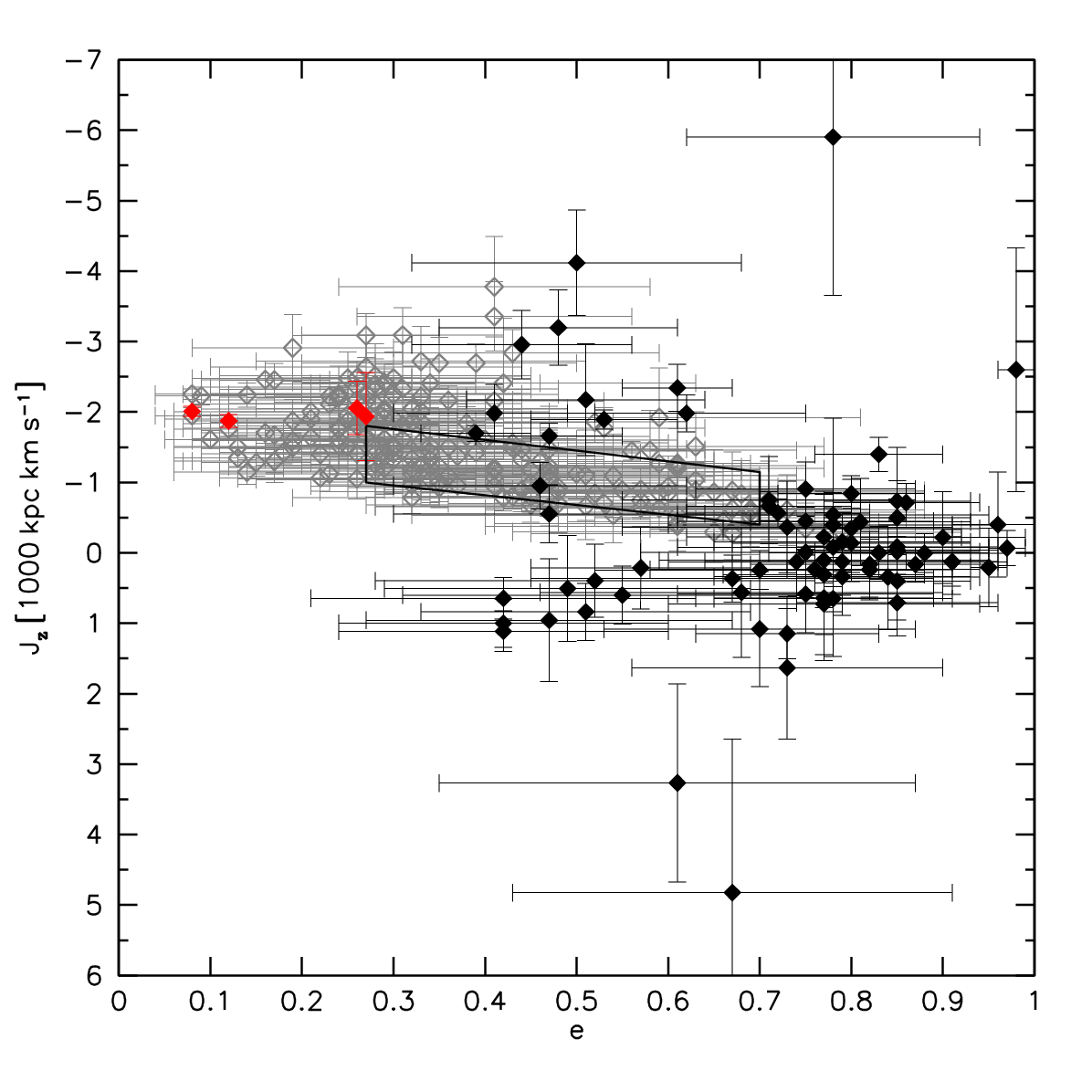}}
\end{center} 
\caption{Left panel: $U-V$ velocity diagram. Red filled diamonds mark thin disk stars, grey open diamonds thick disk stars, and black filled diamond halo objects. The thin and thick disk regions from \citet{pauli06} are plotted as red and black ellipses. Right panel: $J_{\rm z}-e$ diagram of the sample using the same colour coding of the symbols. The box marks the typical region, where thick disk stars are found \citep[Region B,][]{pauli06}. Thin disk stars are found to the left of this region.}
\label{kinematic}
\end{figure*}

\subsection{Sample selection and data analysis}

The kinematical properties of hot subdwarfs are the best diagnostic tool to distinguish between the different populations in our Galaxy. Here we present preliminary results from an ongoing project (Geier et al. in prep.). 

To calculate the Galactic orbits and the kinematic parameters of our stars, we followed the approach outlined in \citep{pauli03,pauli06}. Input parameters are the spectroscopic distances, the radial velocities, and the proper motions. To gather all the input data, we made extensive use of the hot subdwarf catalog.

The sample of 280 sdB and sdOB stars studied here was selected from SDSS Data Release 7 \citep{geier15}. Complementary to the sample RV-variable stars analysed in \citet{geier15}, we selected stars showing no RV-variability on short timescales. To derive the atmospheric parameters, we fitted appropriate model grids for the different sub-classes of sdBs and sdOBs \citep{heber00,otoole06,stroeer07} to the SDSS spectra ($R\simeq1800,\lambda=3800-9200\,{\rm \AA},S/N>30$). A fixed set of spectral lines of hydrogen and helium was fitted by means of chi-squared minimization using the FITSB2 routine \citep{napiwotzki04}. 

Radial velocities of the stars were measured along with the atmospheric parameters by fitting model spectra. Comparing our RV measurements with the RVs of second epoch spectra available in the SDSS and LAMOST archives, we checked $\sim60\%$ of the stars for RV variability. For the moderately RV variable sdBs we found and stars without second epochs we adopted $100\,{\rm km\,s^{-1}}$ as systematic uncertainty, for the constant stars $20\,{\rm km\,s^{-1}}$. Spectroscopic distances to our stars were calculated as described in \citet{ramspeck01}, assuming the canonical mass of $0.47\,{\rm M_{\odot}}$ for the subdwarfs. Input parameters are the spectroscopically determined $T_{\rm eff}$ and $\log{g}$ as well as the extinction-corrected, apparent magnitude in the Johnson V-band \citep[see][for details]{geier17}. 

Because it covers our sample almost completely and it combines only the most recent ground-based catalogues PS1, SDSS and 2MASS with Gaia DR1 we used proper motions from the GPS1 catalogue \citep{tian17}. PPMXL \citep{roeser10} proper motions have been used in the few cases, where no GPS1 data were available.

\subsection{Kinematic parameters}

The routine developed by \citet{pauli03} calculates the trajectories of the stars in the Galactic potential and the kinematic parameters. Uncertainties are determined with a Monte Carlo approach. Besides the Galactic velocity components $U$ ($v_{\rm \rho}$), $V$ ($v_{\rm \theta}$), $W$, the most important parameters to distinguish between the Galactic populations are the eccentricity of the orbit $e$ and the component of the orbital angular momentum perpendicular to the Galactic plane $J_{\rm z}$. 

The kinematic parameters are used to determine the most likely Galactic population for all the stars in our sample. The most direct constraint is provided by the $z$ distance from the Galactic plane. Since the scale-height of the thin disk is just about $0.1\,{\rm kpc}$ and the one of the thick disk about $3.5\,{\rm kpc}$ above and below the Galactic plane, we conclude that objects with $|z|>0.5\,{\rm kpc}$ must be either thick disk or halo stars whereas stars with $|z|>4.0\,{\rm kpc}$ must belong to the halo population. In this way we identify 17 halo stars. The other objects are too close to the Galactic plane and the kinematical properties have to be used to separate the populations.

We used the criteria determined by \citet{pauli06}, who defined specific regions in the $U-V$ and the $J_{\rm z}-e$ diagrams as characteristic for the thin and thick disk populations. Objects outside of this regions most likely belong to the Galactic halo (see Fig.~\ref{kinematic}). 

\subsection{Comparison of populations}

The majority of the stars are members of the thick disk. Only four objects fufill the criteria of being thin disk stars. However, we identified a significant sample of 78 halo sdB star candidates. To select a halo sample that is as pure as possible, we also singled out 57 stars with kinematic parameters and $z$-distances that do not overlap with the thick disk region within their uncertainties. This subsample was called the bona fide halo sample.

\begin{figure*}[t!]
\begin{center}
	\resizebox{8cm}{!}{\includegraphics{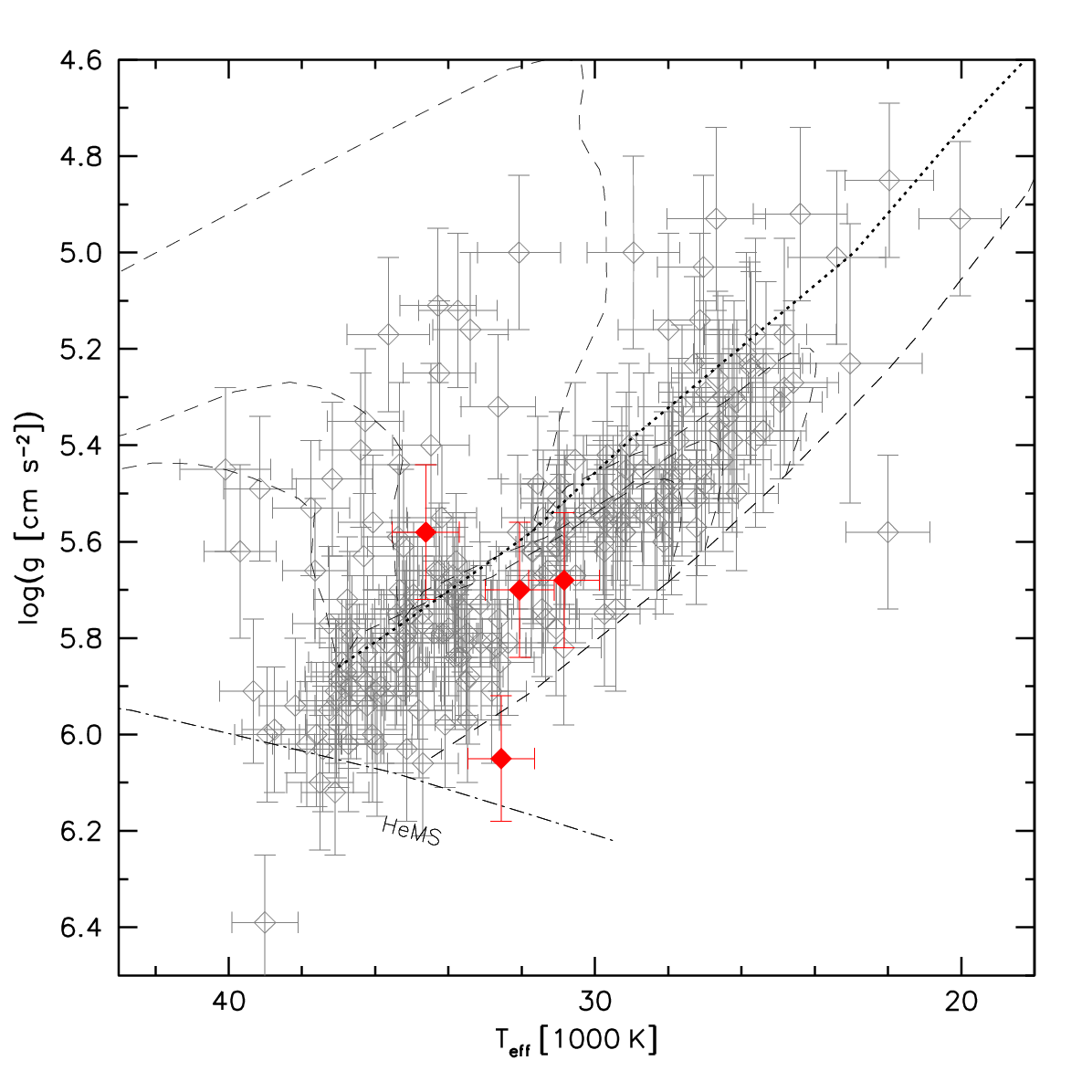}}
	\resizebox{8cm}{!}{\includegraphics{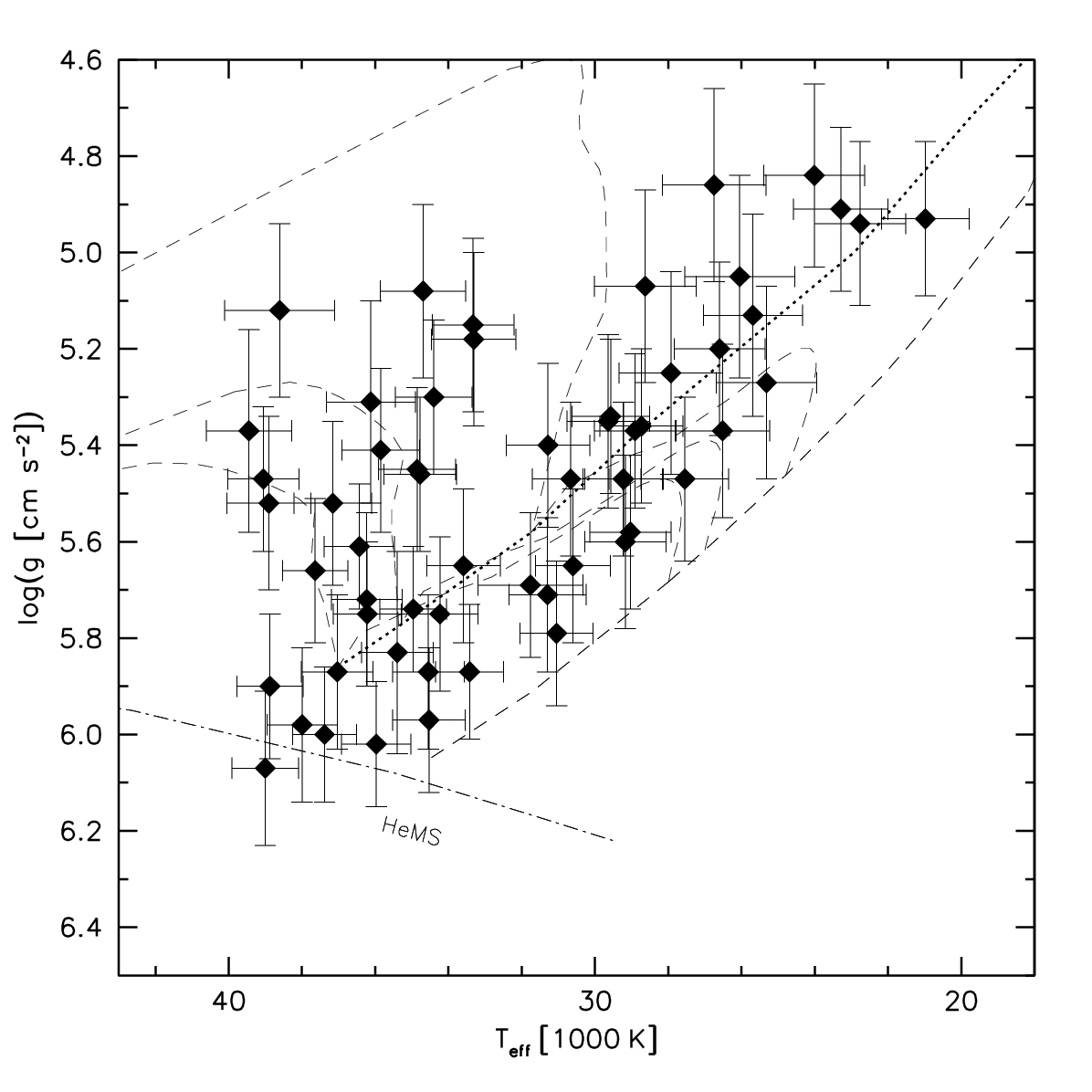}}
\end{center} 
\caption{$T_{\rm eff}-\log{g}$ diagrams of our sample (red filled diamonds: thin disk, grey open diamonds: thick disk, black filled diamonds: bona fide halo). The helium main sequence (HeMS) and the HB band are superimposed with HB evolutionary tracks (dashed lines) for subsolar metallicity ($\log{z}=-1.48$) from \citet{dorman93}. The three tracks correspond to helium core masses of $0.488$, $0.490$ and $0.495\,M_{\rm \odot}$ (from bottom-left to top-right). The disk sample is shown on the left panel, while the right-hand panel shows the bona fide halo sample.}
\label{tefflogg}
\end{figure*}

Fig.~\ref{tefflogg} shows the $T_{\rm eff}-\log{g}$ of the disk (left panel) and bona fide halo (right panel) samples overplotted with evolutionary tracks of subsolar metallicity $\log{z}=-1.48$ \citep{dorman93}. This metallicity is typical for the old thick disk population and matches the observed distribution of atmospheric parameters of the disk stars quite well. The EHB is densely populated and quite well defined. The post-EHB region is also populated, but with a much lower density as expected from the evolutionary timescales. 

For the halo population the distribution is systematically shifted to the upper left. The EHB defined by the moderately subsolar metallicity evolutionary tracks does not match the observed distribution of halo stars as well as the disk stars. This indicates a population of even lower metallicity that is typically observed in the halo \citep{xiong17}. 
                                     
The field halo population can now be compared with the hot subdwarfs discovered in globular clusters, which also belong to the halo and should be of similar age. \citet{latour14} analysed 32 hot subdwarfs of sdB and sdOB type in $\omega$~Cen and found that only $22\%$ of this sample were hydrogen-rich sdB stars, which constitute the majority of hot subdwarfs in all the field samples. The majority of $78\%$ are sdOB stars with temperature around $35\,000\,{\rm K}$ and a moderate enrichment of helium above the solar value ($\log{n({\rm He})/n({\rm H})}=-1.0..0.0$). \citet{moehler04} found 9 sdOBs with supersolar He abundance out of 19 stars ($47\%$) in NGC\,2808. 

\begin{figure*}[t!]
\begin{center}
	\resizebox{8cm}{!}{\includegraphics{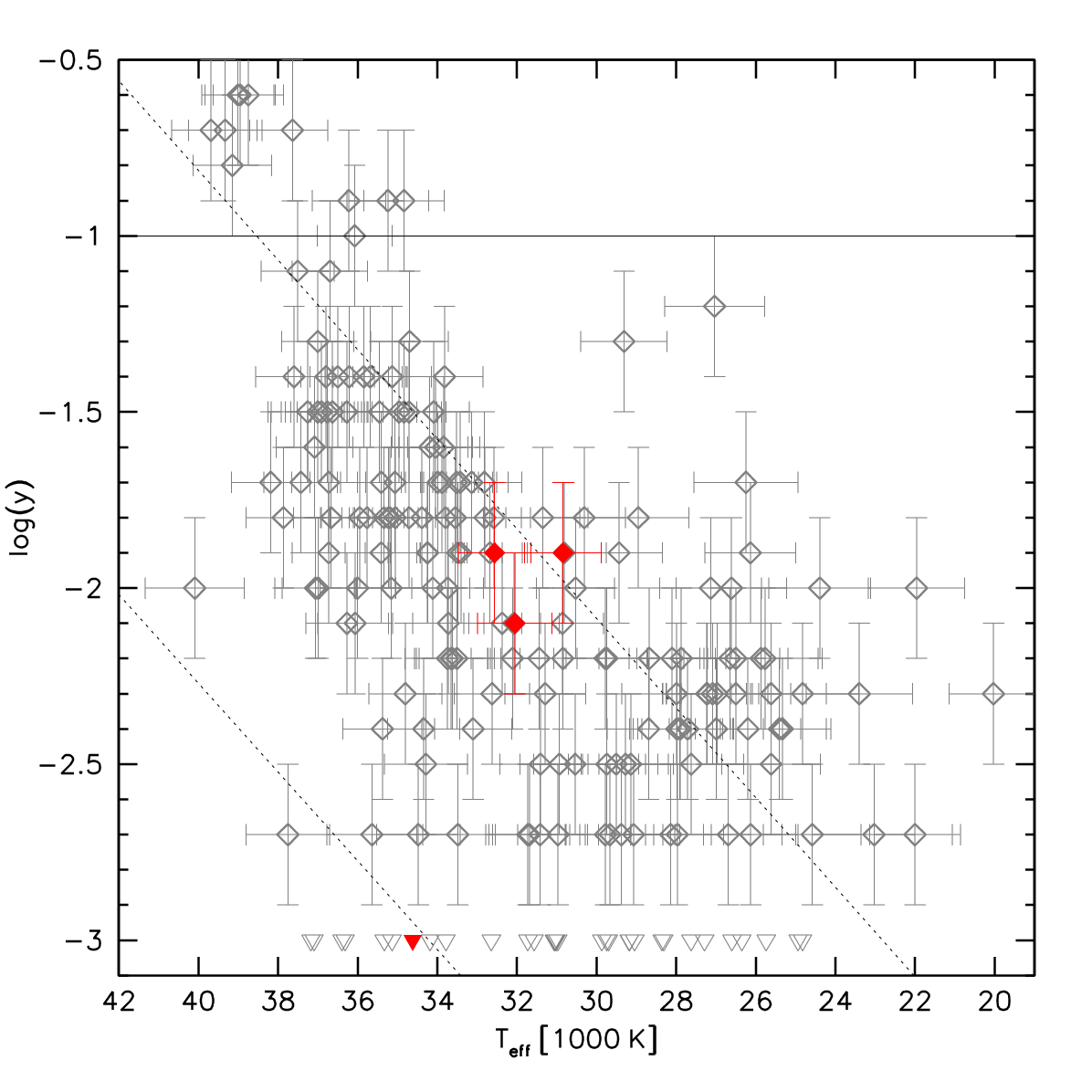}}
	\resizebox{8cm}{!}{\includegraphics{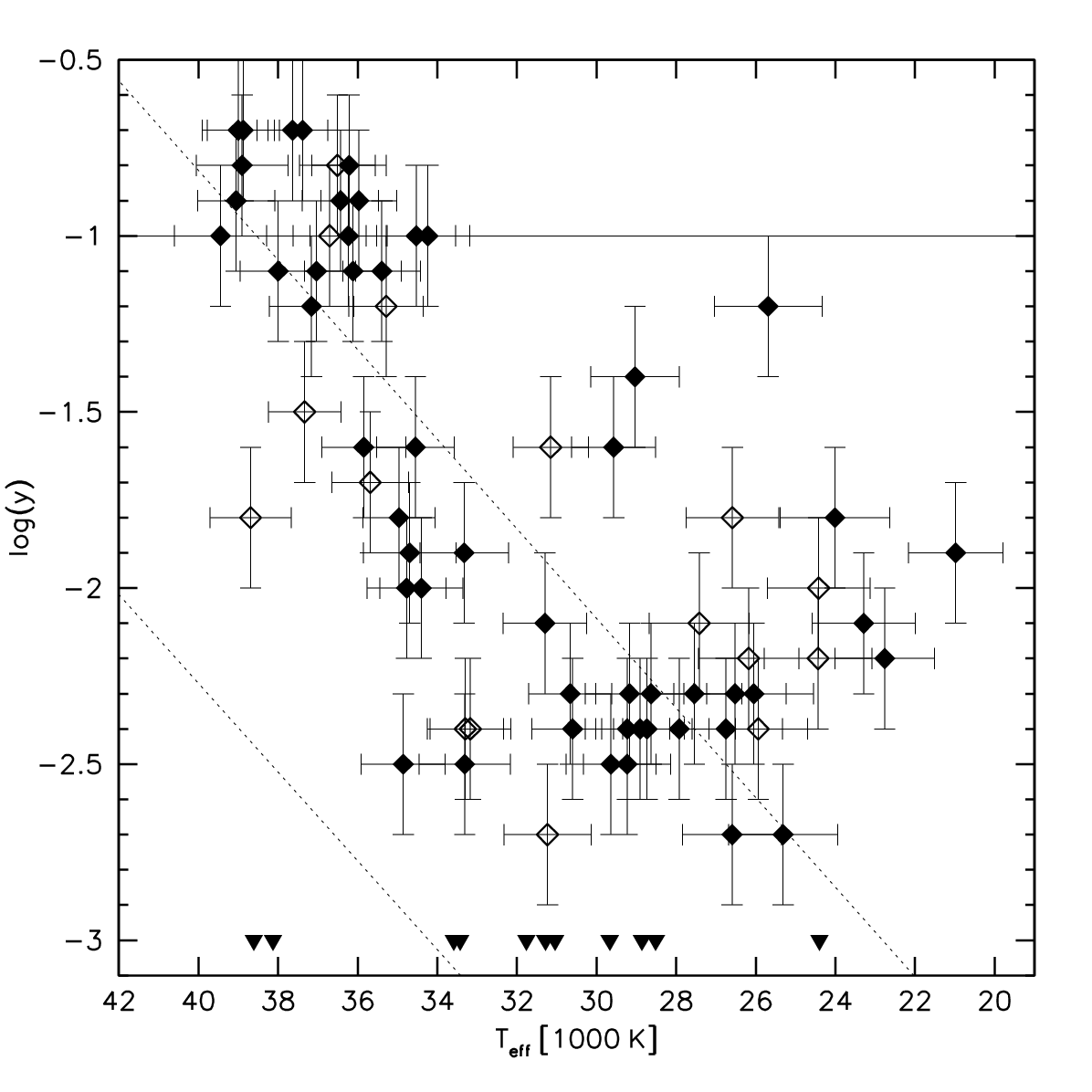}}
\end{center} 
\caption{Helium abundance plotted against effective temperature (red filled diamonds: thin disk, grey open diamonds: thick disk, black open diamonds: halo, black filled diamonds: bona fide halo, triangles: upper limits). The horizontal lines mark the solar helium abundance. The two helium sequences discovered by \citet{edelmann03} are indicated as dotted lines. The left panel shows the disk sample, right panel shows the halo sample.}
\label{nhevsteff}
\end{figure*}

The bona fide halo sample also contains 13 such objects (see Fig.~\ref{nhevsteff} right panel). However, their fraction is only as high as $23\%$. Although this fraction is significantly higher than in the disk sample, where only 11 out of 206 stars or $5\%$ are intermediate He-sdOBs, it is still much smaller than in the globular cluster samples.

\section{Conclusion}

The new catalog of hot subdwarfs contains a significant fraction of the total sdO/B population from the thin disk, thick disk, and the halo. This sample will be cross-matched with the Gaia catalog and used to define the criteria (reduced proper motions, distances, color cuts in the Gaia bands, etc.) for the selection of a homogeneous all-sky catalog of sdO/B stars. 

A full kinematic analysis of a sample of 280 stars was performed using this catalog. This allowed us to separate the first significantly large sample of 78 sdBs and sdOBs belonging to the Galactic halo. Comparing the properties of hot subdwarfs from the disk and the halo with hot subdwarf samples from the GCs $\omega$~Cen and NGC\,2808, we found the fraction of intermediate He-sdOBs in the field halo population to be significantly lower than in the GCs.


\begin{thebibliography}{} 

\bibitem[Altmann et al., 2004]{altmann04}
Altmann, M., Edelmann, H., \& de Boer, K. S. 2004, A\&A, 414, 181

\bibitem[Altmann et al., 2017]{altmann17}
Altmann, M., Roeser, S., Demleitner, M., Bastian, U., \& Schilbach, E. 2017, A\&A, 600, 4

\bibitem[de Boer et al., 1997]{deboer97}
de Boer, K. S., Aguilar Sanchez, Y., Altmann, M., et al. 1997, A\&A, 327, 577

\bibitem[Dorman et al., 1993]{dorman93}
Dorman, B., Rood, R. T., \& O'Connell, R. W. 1993, ApJ, 419, 596

\bibitem[Edelmann et al., 2003]{edelmann03}
Edelmann, H., Heber, U., Hagen, H.-J., et al. 2003, A\&A, 400, 939

\bibitem[Geier et al., 2015]{geier15}
Geier, S., Kupfer, T., Heber, U., et al. 2015b, A\&A, 577, 26

\bibitem[Geier et al., 2017]{geier17}
Geier, S., \O stensen, R. H., Nemeth, P., et al. 2017, A\&A, 600, 50

\bibitem[Heber, 1986]{heber86}
Heber, U. 1986, A\&A, 155, 33

\bibitem[Heber, 2016]{heber16} 
Heber, U. 2016, PASP, 128, 082001

\bibitem[Heber et al., 2000]{heber00} 
Heber, U., Reid, I. N., \& Werner, K. 2000, A\&A, 363, 198

\bibitem[Humason \& Zwicky, 1947]{humason47}
Humason, M. L., \& Zwicky, F. 1947, ApJ, 105, 85

\bibitem[Kawka et al., 2015]{kawka15}
Kawka, A., Vennes, S., O'Toole, S. J., et al. 2015, MNRAS, 450, 3514

\bibitem[Kepler et al., 2015]{kepler15}
Kepler, S. O., Pelisoli, I., Koester, D., et al. 2015, MNRAS, 446, 4078

\bibitem[Kepler et al., 2016]{kepler16}
Kepler, S. O., Pelisoli, I., Koester, D., et al. 2016, MNRAS, 455, 3413

\bibitem[Kilkenny et al., 1988]{kilkenny88}
Kilkenny, D., Heber, U., \& Drilling, J. S. 1988, SAAOC, 12, 1

\bibitem[Kilkenny et al., 2016]{kilkenny16}
Kilkenny, D., Worters, H. L., O'Donoghue, D., et al. 2016, MNRAS, 459, 4343

\bibitem[Latour et al., 2014]{latour14}
Latour, M., Randall, S. K., Fontaine, G., et al. 2014, ApJ, 795, 106

\bibitem[Martin et al., 2017]{martin17}
Martin, P., Jeffery, C. S., Naslim, N., \& Woolf, V. M. 2017, MNRAS, 467, 68

\bibitem[Moehler et al., 2004]{moehler04}
Moehler, S., Sweigart, A. V., Landsman, W. B., Hammer, N. J., \& Dreizler, S. 2004, A\&A, 415, 313

\bibitem[Napiwotzki et al., 2004]{napiwotzki04} 
Napiwotzki, R., Yungelson, L., Nelemans, G. et al. 2004, ASP Conf. Ser., 318, 402

\bibitem[\O stensen, 2006]{oestensen06}
\O stensen, R. H., 2006, Baltic Astronomy, 15, 85

\bibitem[O'Toole \& Heber, 2006]{otoole06}
O'Toole, S. J., \& Heber, U. 2006, A\&A, 452, 579

\bibitem[Pauli et al., 2003]{pauli03}
Pauli, E.-M., Napiwotzki, R., Altmann, M., et al. 2003, A\&A, 400, 877

\bibitem[Pauli et al., 2006]{pauli06}
Pauli, E.-M., Napiwotzki, R., Heber, U, Altmann, M., \& Odenkirchen, M. 2006, A\&A, 447, 173

\bibitem[Ramspeck et al., 2001]{ramspeck01}
Ramspeck, M., Heber, U., \& Edelmann, H. 2001, A\&A, 379, 235

\bibitem[Roeser et al., 2010]{roeser10}
Roeser, S., Demleitner, M., \& Schilbach, E. 2010, AJ, 139, 2440

\bibitem[Str\"oer et al., 2007]{stroeer07}
Str\"oer, A., Heber, U., Lisker, T., et al. 2007, A\&A, 462, 269

\bibitem[Tian et al., 2017]{tian17}
Tian, H.-J., Gupta, P., Sesar, B., et al. 2017, ApJ, submitted (arXiv:1703.06278v2)

\bibitem[Vennes et al., 2011]{vennes11} 
Vennes, S., Kawka, A., \& N\'emeth, P. 2011, MNRAS, 410, 2095

\bibitem[Xiong et al., 2017]{xiong17}
Xiong, H., Chen, X., Podsiadlowski, P., et al. 2017, A\&A, 599, 54

\bibitem[Wisotzki et al., 1996]{wisotzki96}
Wisotzki, L., Koehler, T., Groote, D., \& Reimers, D. 1996, A\&AS, 115, 227

\bibitem[Zacharias et al., 2017]{zacharias17}
Zacharias, N., Finch, C., \& Frouard, J. 2017, AJ, 163, 166

\end{thebibliography}
\end{document}